\documentclass[sigconf, screen]{acmart}
\usepackage{blindtext}

\usepackage{mathrsfs}

\usepackage{soul}

\usepackage{etoolbox}
\makeatletter
\patchcmd\@thm
 {\let\thm@indent\noindent}%
  {}{}
\makeatother

\usepackage{dsfont}
\usepackage{enumitem}

\renewcommand\footnotetextcopyrightpermission[1]{} 

\usepackage{booktabs} 
\usepackage{mdframed}
\usepackage{framed}

\usepackage{hyphenat}
\usepackage{fnpct}

\usepackage{algpseudocode}
\usepackage{amsmath,epsfig}
\usepackage[boxed,ruled,linesnumbered, longend, noend]{algorithm2e}
\usepackage{amsthm}
\usepackage{setspace}
\usepackage{enumitem}
\usepackage{multirow}
\usepackage{hhline}
\usepackage{caption}
\usepackage{subcaption}
\usepackage{graphicx}
\usepackage{wrapfig}
\usepackage{amsthm}

\setcopyright{rightsretained}
\usepackage{tikz}
\usepackage{balance}

\usepackage{csquotes}
\usepackage{enumitem}
\usepackage{tablefootnote}
\usepackage{fnpct}
\usepackage{bbm}
\usepackage{lipsum}
\usepackage{setspace}

\makeatletter
\def\thm@space@setup{\thm@preskip=0pt
\thm@postskip=0pt}
\makeatother

\usepackage{array}
\usepackage{makecell}

\newcommand{\B}{\vspace*{-\smallskipamount}}
\newcommand{\BB}{\vspace*{-\medskipamount}}
\newcommand{\BBB}{\vspace*{-\bigskipamount}}

\usepackage{longtable}
\usepackage{mdframed}
\usepackage{colortbl}

\usepackage[hang,flushmargin]{footmisc}

\usepackage{natbib}
\newcommand*\wrapletters[1]{\wr@pletters#1\@nil}
\def\wr@pletters#1#2\@nil{#1\allowbreak\if&#2&\else\wr@pletters#2\@nil\fi}
\usepackage{tikz}
\usetikzlibrary{automata,matrix,shapes,arrows,positioning,chains,calc}
\usetikzlibrary{snakes}
\usetikzlibrary{arrows,scopes}
\usetikzlibrary{positioning,chains,fit,shapes,calc}
\usepackage{caption}
\usepackage{subcaption}
\usepackage{amsmath}
\usepackage{lineno}

\definecolor{lightyellow}{HTML}{ABEBC6}
\sethlcolor{lightyellow}

\definecolor{myblue}{HTML}
{BF40BF}

\mdfdefinestyle{MyFrame}{%
    linecolor=white,
    outerlinewidth=2pt,
    innertopmargin=4pt,
    innerbottommargin=4pt,
    innerrightmargin=4pt,
    innerleftmargin=4pt,
        leftmargin = 4pt,
        rightmargin = 4pt,
        backgroundcolor=yellow!41!white}

\begin{document}

 \newcommand{\sysName}{\textsc{Health$^+$}\xspace}

\newcounter{quescount}
\setcounter{quescount}{1}
\newcommand{\resques}[1]{%
  \vspace{0.1cm}%
  \noindent%
  \colorbox{yellow!20}{
    \parbox{0.95\linewidth}{\textbf{Research Direction \thequescount{}:} \textit{#1}}%
  }%
  \stepcounter{quescount}%
}

\setlength{\belowdisplayskip}{0pt}
\setlength{\belowdisplayshortskip}{0pt}
\setlength{\abovedisplayskip}{0pt}
\setlength{\abovedisplayshortskip}{0pt}


\fancyhead{}

\title{
\textsc{Health}$^+$: Empowering Individuals via Unifying Health Data}

\author{Sujaya Maiyya}
\authornote{\scriptsize This project was made possible in-part through the support of the National Cybersecurity Consortium and the Government of Canada (CSIN) (project id 2024-1345).}
\affiliation{%
  \institution{University of Waterloo}
  \country{Canada}}

\author{Shantanu Sharma}
\affiliation{%
  \institution{New Jersey Institute of Technology}
  \country{USA}}

\author{Avinash Kumar}
\affiliation{
\institution{Independent Researcher}
\country{USA}
}
\authornote{\scriptsize Present affiliation: Google, Mountain View, USA. Work done in personal capacity; views expressed do not represent those of Google. \\
{\color{blue}\textbf{This paper has been accepted in ACM Multimedia 2025.}}}



\begin{abstract}
Managing personal health data is a challenge in today's fragmented and institution-centric healthcare ecosystem. Individuals often lack meaningful control over their medical records, which are scattered across incompatible systems and formats. This vision paper presents \sysName, a user-centric, multimodal health data management system that empowers individuals (including those with limited technical expertise) to upload, query, and share their data across modalities (e.g., text, images, reports). Rather than aiming for institutional overhaul, \sysName emphasizes individual agency by providing intuitive interfaces and intelligent recommendations for data access and sharing. At the system level, it tackles the complexity of storing, integrating, and securing heterogeneous health records, ensuring both efficiency and privacy. By unifying multimodal data and prioritizing patients, \sysName lays the foundation for a more connected, interpretable, and user-controlled health information ecosystem.
\end{abstract}



\maketitle

\BB
\section{Introduction}
\label{sec:intro}
\textit{Have you ever faced an ailment such as back pain, sprained ankle, or fatigue? If so, how many doctors and medical professionals did you interact with for an accurate diagnosis and treatment? And how easy was it to share your medical records with these professionals?}

In today's healthcare landscape, patient data spans a rich spectrum of multimedia content: diagnostic images (e.g., MRIs, X-rays), clinical audio and video consultations, PDF reports, time-series signals from wearable devices, and structured data from electronic health records (EHRs). However, this wealth of information remains disjointed across siloed systems, each managed independently by hospitals, imaging centers, laboratories, and personal health apps.

 While the vision of a unified national healthcare system remains idealistic, the reality is a fragmented ecosystem with heterogeneous data modalities, inconsistent data-sharing policies, and limited patient access.  For example, in North America, which is the third-largest continent with 592 million people, there exists no single framework or system
that enables patients to access their collective medical records and to share them with other medical professionals. 
 In the US, the widely used EHR system Epic~\cite{epic} is deployed in only 37\% of hospitals~\cite{epic-usage}. Similarly, in India, the world's most populous country with 1.4 billion residents, no unified framework exists even at the state or district level.

 This fragmentation not only hinders clinicians from accessing comprehensive patient histories~\cite{idowu2023streams}, but also leaves individuals without a reliable way to collect, query, or control their health data. 

This vision paper aims to explore challenges and design novel techniques to develop a holistic framework, entitled \textbf{\sysName, a user-centric health data platform that unifies and empowers access to an individual's entire medical history by treating it as a multimedia corpus of sensitive personal data}. \sysName will facilitate multimodal ingestion, storage, fusion, querying, and sharing of data, with a strong emphasis on \emph{usability and privacy}.

In an ideal world, all hospitals and healthcare providers in a country use the same system to store and manage patient data, and the patients have complete access to any of their medical information. However, striving for this ideal scenario is impractical since requiring to convince \textit{all} existing healthcare providers to switch to a new system, and likely involving legal and legislative changes. 
We deem this solution implausible, and hence, take a radically different approach in which the healthcare providers will continue to use their existing systems with no changes, while the onus of 
collecting various pieces of medical data and providing it to a unified health management system, such as \sysName lies with the patients. The challenges involved in transfer and management of such data is discussed later.
We believe that enabling patients to consolidate their 
medical records in one place \textit{today} at the expense of requiring a proactive patient takes precedence over striving for an ideal solution. 

To ascertain the relevance of such a unified and patient/user-centric health management system, we present an example and use it as a running example throughout the paper. 
The motivation for this vision paper stems from repeated difficulties individuals experience in navigating medical systems; the following example conveys these difficulties in a consolidated way.

\BB
\subsection{Motivating Use-Case: Lower Back Pain}
\B
\noindent
\textbf{Introducing disc herniation.}
Lower back pain affects $\approx$80\% of the population at least once in their lives~\cite{freburger2009rising,wheeler2016evaluation,who}. Among the various reasons for this discomfort, intervertebral degeneration stands out as the most common, often resulting in lumbar disc herniation or degenerative disc disease. The spinal column in the human body consists of 23 vertebral discs, each consisting of two components: a resilient ring-shaped outer layer known as the \emph{annulus fibrosus}, and a soft, gel-like inner structure known as the \emph{nucleus pulposus}.

A herniated disc is characterized by damage to the annulus fibrosus, allowing the nucleus pulposus to herniate (rupture or protrude), leading to symptoms such as lower back pain, fatigue, tingling sensations, pain in one or both legs, and difficulty sitting. The problem of disc herniation can be identified through magnetic resonance imaging (MRI) scans and typically necessitates either conservative medical treatments like physical therapy or surgical interventions. Annually, the economic burden of lower back pain surpasses \$100 billion in the United States~\cite{katz2006lumbar,al2020lumbar}!

\smallskip
\noindent
\textbf{Complexity of data management and absence of a unified view.}
Let us delve into the intricacies involved in treating a herniated disc and the associated challenges in data management and sharing. The treatment process comprises multiple entities, including a primary care physician, imaging centers employing X-ray and MRI technologies, and a specialized team featuring a pain specialist, physical therapist, orthopedic surgeon, and neurosurgeon.

\begin{figure}[!t]
\BBB\BBB\BBB\BB
    \centering
    \includegraphics[scale=0.33]{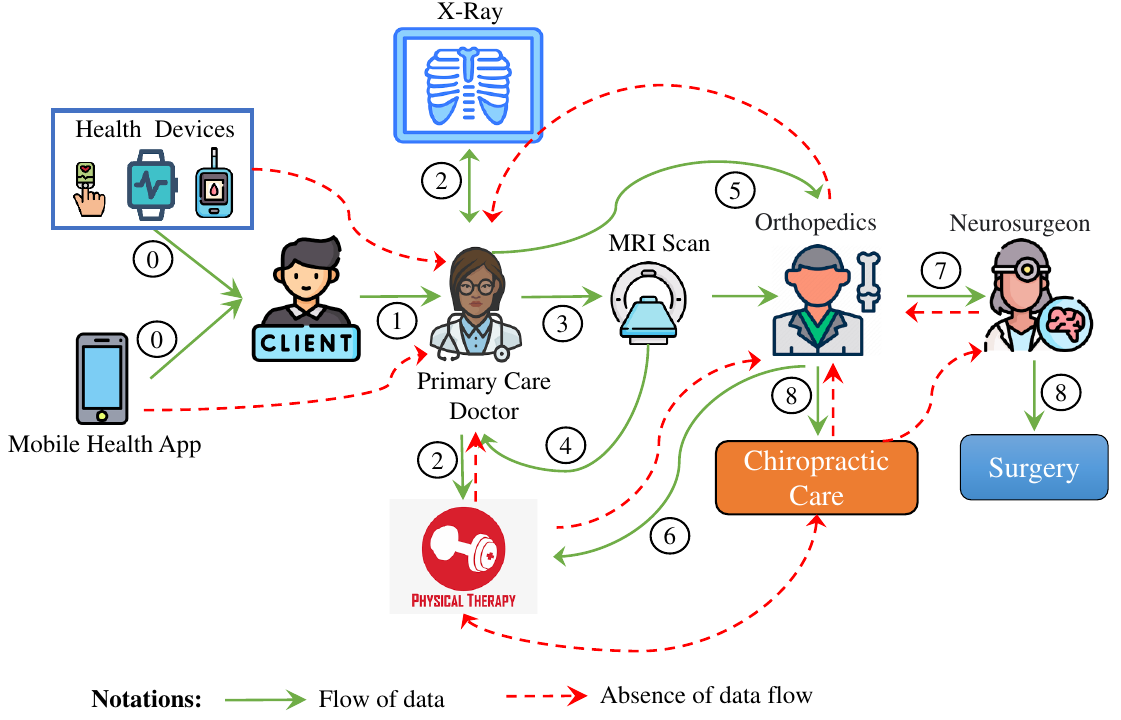}
    \BBB\BB
    \caption{Disc herniation treatment process.}
    \label{fig:Disc herniation treatment process in the US}
    \BBB\BB
\end{figure}

{\color{blue}Figure~\ref{fig:Disc herniation treatment process in the US}} shows the treatment process and the flow of data among different entities. The treatment begins with a primary care physician, who may initially suggest X-ray assessments at imaging centers and physical therapy sessions to alleviate the symptoms. After several weeks, if the pain persists, the primary care doctor may prescribe an MRI, which yields a CD Disc along with a detailed written report.\footnote{\scriptsize In the US, immediate MRI procedures (in the first week of the injury) may be subject to insurance regulations that necessitate prior completion of physical therapy.} Typically, the locations where imaging procedures are conducted are distinct from the clinics where primary care physicians operate. Hence, the MRI report must be propagated back to the primary care doctor, who, based on the findings, might refer the patient to an orthopedic surgeon. The orthopedic surgeon may further recommend visits to a physical therapist, chiropractic care and, in some cases, refer the patient to a neurosurgeon for specialized care.

The various healthcare professionals mentioned above utilize different online portal systems~\cite{idowu2023streams} or rely solely on paper-based methods. 
This places an additional burden on patients, who must carry all referrals and their medical history when scheduling appointments with doctors or specialists, and thus, impede individuals from changing doctors as needed or seeking second opinions. Additionally, the appointment scheduling process involves complex questions with legal and medical language, which patients often fail to comprehend. For example, forms necessary to book appointments can include a consent section, requesting permission to transfer an individual's data to another doctor or research organization upon request. Lacking an understanding of the implications and repercussions of such forms, patients often either accept or deny \textit{all} such requests, with denials posing additional challenges in transferring their records to another doctor. In such instances, patients may be required to pay several hundred dollars to obtain a physical copy of their records! Doctors take a long time to ascertain the exact cause of the problem, as they lack immediate access to a patient's past medical history and information about their daily life.

\medskip
\noindent
\textbf{Reasons for complexities.} The primary sources of such complications in the medical systems are:

\begin{figure}[!t]
\BBB\BBB\BBB\BB
\centering
\begin{subfigure}{.49\linewidth}
\hspace{-2.5cm}  
\centering
  \includegraphics[scale=0.15]{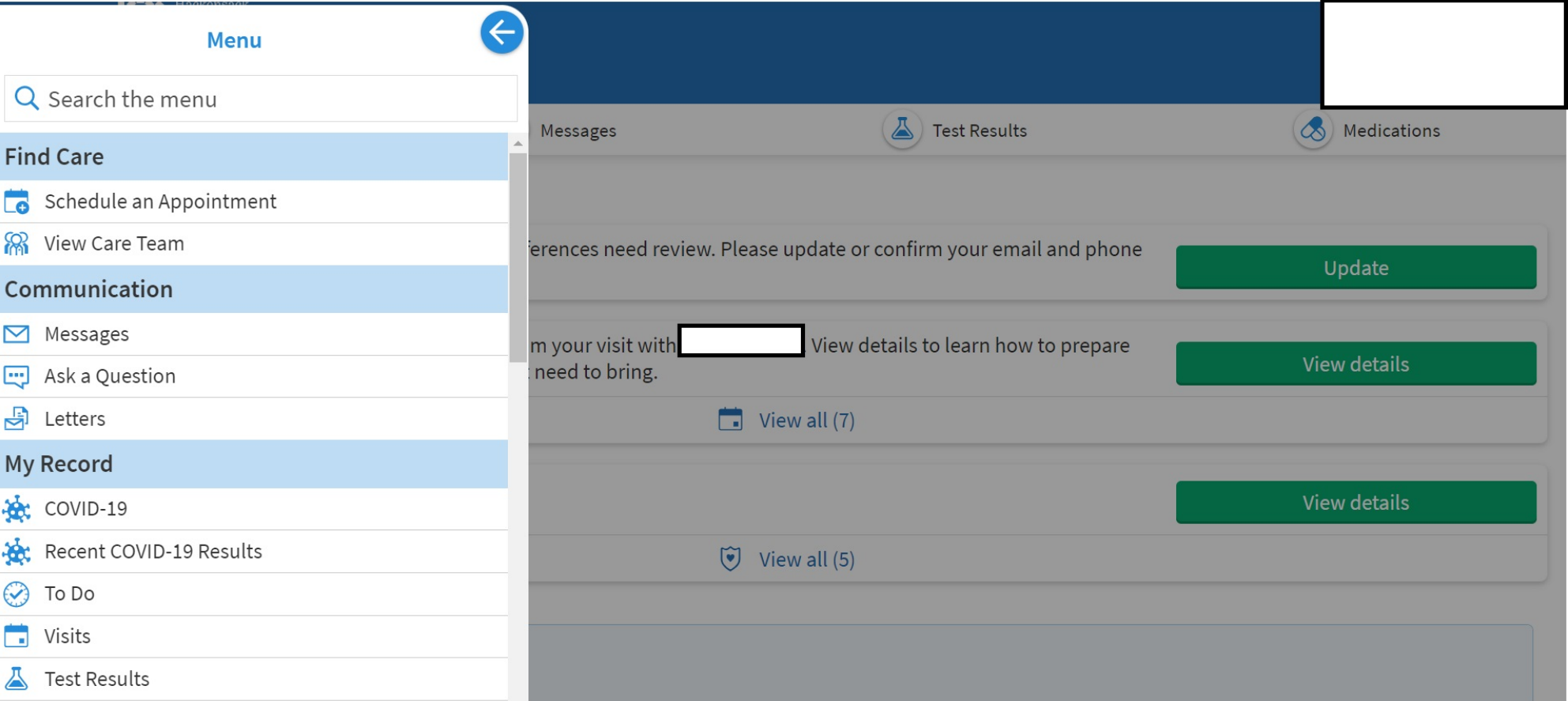}
  \label{fig:sub1}
\end{subfigure}%
\begin{subfigure}{.49\linewidth}
  \centering
  \includegraphics[scale=0.15]{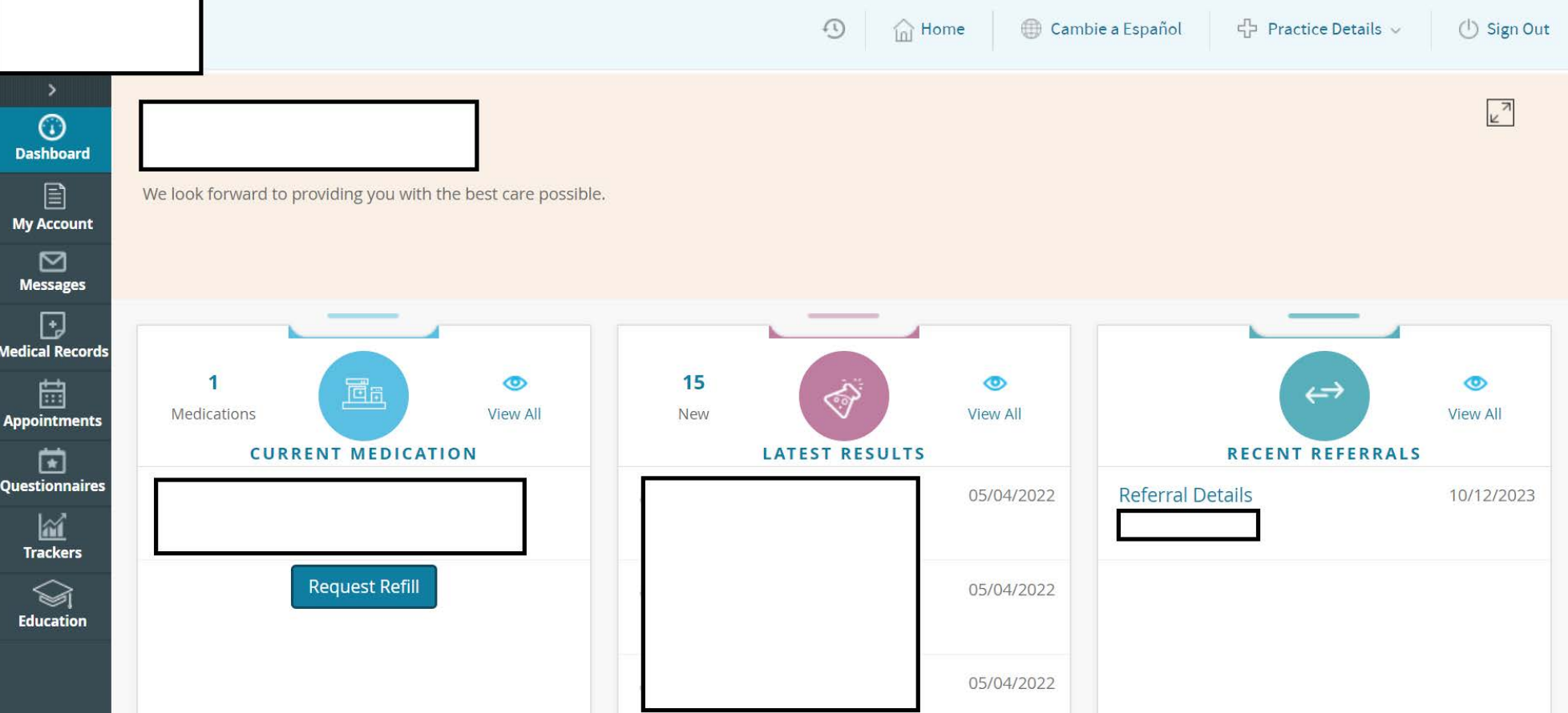}
  \label{fig:sub2}
\end{subfigure}
\BBB\B
\caption{\centering Views of two different frameworks an individual needs to access for visiting two different healthcare providers}
\label{fig:twoviews}
\BBB\BB
\end{figure}
\begin{enumerate}[noitemsep,nolistsep,
leftmargin=0.01in]

\item \textbf{\emph{Diverse Systems, Modalities, and Stakeholders.}} 
Health data is generated and consumed by a wide array of actors --- patients, doctors, hospitals, labs, insurers, and wellness apps --- each using distinct and incompatible systems and pursuing different goals.
These systems capture data across multiple modalities, such as structured medical records, time-series sensor data, and unstructured media like radiology images or consultation audio, and data flows unidirectionally, as depicted in~{\color{blue}Figure~\ref{fig:Disc herniation treatment process in the US}}. These diverse actors operate independently, creating an isolated and incomplete experience for the patient by allowing them to access only a subset of their multimodal health data.

\item
\textbf{\emph{Lack of User-Centric Privacy in Multimodal  Systems.}} 
Privacy in healthcare systems is frequently enforced at the level of institutions rather than individuals, leading to a critical paradox: while privacy laws (e.g., HIPPA and GDPR) aim to protect user privacy, their current implementations often deny users access to their data.
This issue becomes especially problematic in multimodal health data, where each modality is typically governed by separate systems, platforms, and vendors, resulting in fragmented privacy boundaries. In practice, privacy is often invoked as a barrier: individuals may need to pay hundreds of dollars, wait weeks, or navigate complex interfaces to access even basic records. Worse, they are rarely empowered to curate, annotate, or integrate multimodal streams (e.g., combining MRI scans, physical activity from wearables, and symptom logs) to gain personal insights or to share with new providers.

\item \textbf{\emph{Fragmented Clinical Views.}} 
Clinicians rarely have access to a holistic view of a patient's health trajectory. When health records are fragmented across incompatible systems, it becomes difficult to correlate MRI scans, doctor’s notes, audio consultation logs, and fitness tracker trends to form a coherent diagnosis. As a result, key temporal and semantic connections across modalities are lost. Doctors often must rely on patient memory for historical or behavioral context, yet patients are not trained to surface all medically relevant details. Meanwhile, intelligent digital traces from wearables and apps remain underutilized, despite their potential to explain chronic or lifestyle-based conditions~\cite{bourgeois2010patients}.

\item
\textbf{\emph{Absence of Multimodal Integration and Reasoning Capabilities.}} 
Current health data infrastructures lack the capability to integrate, represent, and reason across multiple data modalities, which exist in silos, each with their own storage, access protocols, and semantic structures. While the multimedia community has made significant progress in aligning and fusing multimodal data streams for tasks such as sentiment analysis, video understanding, and cross-modal retrieval, such techniques are seldom applied in health contexts, which is highly privacy sensitive. As a result, systems fail to answer fundamental questions that require cross-modal inference, e.g., correlating heart rate variability and self-reported symptoms to personalize diagnosis or treatment.

\end{enumerate}

In summary, modern healthcare suffers from a critical lack of multimedia integration, continuity, and user-centric control.

\subsection{Our Objectives}
\B
The goal of \sysName is to directly address these issues by treating personal health information as a multimedia data problem, empowering users with more control over their medical data and providing a single solution that collectively stores all of user's medical history. Materializing this objective encompasses various steps, which we describe from two perspectives: a user's or an individual's perspective and a system perspective.

\begin{enumerate}[noitemsep,nolistsep,leftmargin=0.01in]
    \item \textbf{\emph{\textit{Individuals' perspective.}}} We aim to design \sysName in a way that enables even technologically challenged users to easily interact with the system by uploading, querying, and downloading their medical records. 
Beyond basic access, \sysName leverages multimodal reasoning to autonomously recommend which subset of their records should be shared with different types of healthcare professionals without requiring domain expertise. Furthermore, \sysName analyzes and detects anomalies (e.g., missing or inconsistent entries) within the user-provided data across modalities, offering proactive feedback for improving data quality.

    \item \textbf{\emph{\textit{System perspective.}}} With regard to designing a unified health system, \sysName aims to contribute the following: 
    (i) Efficient storage and retrieval: to support storage of heterogeneous media types with low-latency retrieval.
(ii) Cross-modal fusion for semantic querying:  to resolve semantically complex user queries (e.g., integrating MRI scan metadata, pain progression logs, and physician notes for disc herniation).
(iii) Privacy-preserving multimodal data management: to ensure data protection regardless of storage environments. 
    
\end{enumerate}

\section{Why Now?}
\BB
Recent years have witnessed a surge in multimodal machine learning research applied to healthcare systems --- examples include works on cross-modality image reconstruction from brain activity~\cite{DBLP:conf/mm/XieZZWNX24}, multimodal diagnostic fusion using X-rays, ECGs, and clinical reports~\cite{DBLP:journals/corr/abs-2410-16239}, and retrieval-augmented EHR analysis with large language models~\cite{DBLP:journals/corr/abs-2402-07016}. \cite{DBLP:journals/pvldb/CaiZJOY24,
DBLP:journals/corr/abs-2301-03829,
DBLP:conf/sigmod/Zheng0HNOG21,
DBLP:journals/pvldb/ZhengCCHZO22} have focused on health data analytics and health prediction. For example,~\cite{DBLP:journals/pvldb/CaiZJOY24} focuses on finding cohorts, 
\cite{DBLP:journals/pvldb/ZhengCCHZO22} for dynamic healthcare analytics,
\cite{DBLP:conf/sigmod/Zheng0HNOG21} for medical task decomposition for better prediction, and 
\cite{DBLP:journals/corr/abs-2301-03829} for predicting nutrition in food.
 Complementary efforts have also explored risk prediction~\cite{DBLP:conf/mm/SawhneyMMKSZ20}, sentiment analysis across modalities~\cite{DBLP:conf/mm/HuangJYS23}, and review works discussing health data~\cite{DBLP:journals/hij/PengGB20,info:doi/10.2196/jmir.5094}.

While these efforts demonstrate the feasibility and power of multimodal learning, they remain narrowly scoped, being either task-specific, dataset-bound, and/or institution-centered, which limits their applicability for everyday users.
Critically, \emph{\textbf{the existing methods neglect two urgent concerns: (i) empowerment of non-technical end-users by providing control over their data, and (ii) privacy-preserving data access.}}

As health data becomes increasingly multimodal and distributed, there is a pressing need for systems that do not just fuse data across modalities, but also provide intuitive, policy-aware, and privacy-conscious interfaces for patients.

\sysName responds to this moment by rethinking health data interaction from the ground up. First, it assumes users are non-technical individuals who seek to navigate complex health interactions without expertise in data science. Unlike data lake solutions designed for technical analysts~\cite{conf/dawak/GieblerGHSM19, fang2015managing, mathis2017data, conf/medes/MaderaL16}, \sysName offers intelligent assistance to automate tasks like suggesting what data to share with specific providers. Second, it recognizes that health data carries privacy sensitivities far beyond traditional enterprise data. By combining secure multimodal storage with user-facing privacy policies and explainable sharing mechanisms, \sysName bridges the gap between modern multimodal learning and real-world patient empowerment.

\section{Research Directions}
\label{sec:research-dirs}
\B
Before proposing the architecture of \sysName, we highlight the novel research directions that this system has the potential to spawn. \sysName is not merely a data storage platform, it represents a paradigm shift in how multimodal health data can be managed, interpreted, and accessed by non-technical users. From an individual's perspective, it raises new challenges in human-centric interaction with heterogeneous data streams. From a system design perspective, it opens up fertile ground for developing intelligent, privacy-preserving infrastructures that enable real-time fusion, querying, and personalization over distributed multimodal sources.

\subsection{Individuals’ Perspective}
\B
This section discusses the potential research directions spawned from the perspective of how an individual can use a unified health data management system. The terms individual, user, or patient all refer to the person using the system \sysName.

\medskip
\noindent\textbf{System usability.}

Having convinced a reader of the necessity for a proactive patient involved in collecting and uploading their medical data, the first aspect \sysName needs to address is the ease of usability of the system. We envision primarily providing two types of interfaces that patients can use to interact with \sysName: a mobile and a web application. Users must be able to upload electronic health records received from various specialists they interact with or upload pictures of hand-written records (which is still a common practice, especially when interacting with specialists such as physical therapists). Note that we assume a typical user of systems, such as \sysName, to be non-expert both from the medical and from the technological perspectives. Hence, the system should \textit{not} expect the users to clearly classify the data they upload. Once uploaded, users can request to access records using multiple filters, e.g., records from a specific doctor or a hospital, records pertaining to a specific treatment or illness, records from a certain time interval, and so on~\cite{quamar2022natural}. They may also want to request aggregated statistics on the filtered records (e.g., minimum weight or average heart rate during a treatment period). Note that {\color{blue}\S\ref{sub:sys_perspective}} will discuss the system challenges to achieve these functionalities and to ensure privacy and security of stored data.

Enabling an end user (i.e., a patient) to interact with such a system requires research efforts in the User Interface/User Experience (UI/UX) design space. Specific research challenges to focus on involves topics such as supporting different languages requiring accurate translations of complex medical terminologies~\cite{thompson1993doctor}, enabling different audio and visual alternative such that people with disabilities can easily interact with the system, and an intuitive authorization interface to allow authorizing (or revoking) personnel such as primary caregiver to interact with a user's medical data.

\resques{Advancements in UI/UX to enhance the usability of health data management systems}

\medskip
\noindent
\textbf{Enabling data sharing.}

A major functionality of \sysName is enabling a user to share their medical records from one doctor or specialist with other health professionals. The research direction here pertains to how a user can seamlessly share the necessary medical records with a health professional despite having no or limited medical expertise. A trivial solution to achieve this functionality is to share the entire medical history with a professional, whenever some past records are requested. This trivial solution has two challenges, one from a health professional's and one from a patient's standpoint. 

We elaborate on the challenges using an example wherein a patient, Alice, is suffering from two ailments: a disc herniation and a mental disorder of OCD (obsessive compulsive disorder). As part of her disc herniation treatment, Alice has to obtain an MRI scan and interact with a lab technician. The lab technician may request access to certain medical records, such as a physician's diagnosis or recommendation for an MRI scan. If systems, like \sysName, share Alice's \textit{entire} medical history, from the lab technician's standpoint, having to parse through all historical records wastes their precious time and resources. From Alice's perspective, sharing the entire history can be an invasion of privacy since the lab technician now has access to records about her OCD ailment~\cite{journals/ijmi/PereraHTFW11}. However, in a different context, e.g., interactions between Alice and her primary physician, sharing her history with OCD might be relevant due to its impact on her overall health and the effects of potential medications.

The primary research challenge from an individual's perspective here involves enabling a user to share \textit{only} the necessary health records with a given health care professional. Since we assume users are non-experts, the onus of this selective sharing lies upon the health system, such as \sysName. This requires the system to first learn associations between ailments and treatments, and then, to identify the relevant records necessary for a given diagnosis~\cite{journals/ijwis/EderS21, conf/dawak/GieblerGHSM19, conf/icde/SancaA23}. 
A robust health system must allow users to enter minimal information, such as `visiting an MRI scanning lab for disc herniation', and expect a recommendation of all, but only the necessary health records to be shared with the lab technician. 
This requires novel research endeavors in machine learning (ML)- or artificial intelligence (AI)-based association learning and recommendation techniques. In particular, we may need to design novel federated learning techniques~\cite{journals/npjdm/RiekeH0MRABGLMO20,dun2023efficient,huang2020personalized} to distributedly train the models on existing associations between what data are relevant to a given diagnosis or treatment, where the model can be trained independently at different hospitals and merged later on for higher accuracy. Then,  this model with \sysName recommends to users on what specific medical records to share for a given diagnosis or treatment.

\resques{ML- and AI-based recommendation techniques to share relevant medical records}

\medskip
\noindent
\textbf{Data enrichment.}

Beyond core functionalities such as uploading, searching, and sharing medical records, a robust health data system like \sysName must also address the pervasive issue of incomplete or inaccurate information. By leveraging multimodal reasoning across structured formats (e.g., lab results, EHR fields) and unstructured content (e.g., clinical notes, scanned documents, images), \sysName can suggest contextually appropriate corrections or impute missing values.

To provide an example scenario, assume that the information about doctor visits is stored in a relational table ({\color{blue}\S\ref{sub:sys_perspective}} discusses data storage in more detail). Say a user, Alice, visits two different doctors, a general physician and an orthopedic, on the same day. Records from her visit to the physician provide details on her vitals, such as height, weight, and blood pressure, each stored as an attribute in the underlying table; whereas her visit to the specialist did not measure the vitals. In such cases, an intelligent health system can not only identify from the documents uploaded by Alice that some values are missing from her second visit, but also infer that these vitals are unlikely to have changed meaningfully in the same day and use this to populate entries for the second visit.

Such data enrichment, which we define as identifying anomalies in the data and proposing meaningful corrections, comes with its challenges because systems, like \sysName, receive and store heterogeneous data collected from multiple sources. Firstly, we need to design algorithms and ML models to process the heterogeneous ingested data, identify anomalies, and provide suggestions with high confidence. While a few recent works leverage generative AI to impute missing data~\cite{conf/icaiic/KimTS20}, these solutions may not trivially extend to medical contexts, requiring us to develop novel algorithms. Secondly, the system needs to decide when to enrich the data. If a system enriches the data on the fly when a user queries anomalous data (e.g., retrieving information about the orthopedic visit with some attributes having NULL values), it can slow the data retrieval process and affect the user experience. In contrast, if the system enriches the data offline, the results of enrichment will occupy further storage, which can be wasteful if the user rarely accesses the enriched data. 

Any means of addressing the above two challenges only form partial solutions: a system must include the user in finalizing the decision on enriched data. Inaccurately populating missing fields can have much more severe consequences for medical purposes than having missing data. Because of this, a correct health data system should interact with a user by first notifying the user of anomalous data, and then, in confirming the changes to fix the anomaly. Moreover, when a user downloads records that are in the aggregated form of heterogeneous data, the records should visibly indicate system-populated data points to bring this to a medical professional's notice.

\resques{Novel algorithms for enriching data and ensuring its transparent visibility in system design}

\subsection{System Perspective}
\label{sub:sys_perspective}
\B
This section discusses how \sysName will push the research forefronts from a system design perspective. In particular, we focus on the novel challenges introduced by storing and merging heterogeneous medical records and addressing the privacy and security challenges incurred due to handling highly sensitive information.

\medskip
\noindent
\textbf{Data storage.}

From a system design perspective, the first two research questions, we need to explore are \textit{where and how to efficiently store a multitude of heterogeneous medical data collected in differing formats?} 

The easier of the two questions is \textit{where} to store the medical data. Two main alternatives for this are to either store the data locally at a user's device, i.e., on mobile phones and laptops, or to store it on the application side, i.e., on the application-owned infrastructure or on infrastructure rented from third-party cloud vendors such as Amazon AWS~\cite{aws} or Google GCP~\cite{gcp}. The first choice of storing all medical records on a user's device is impractical due to its lack of scalability, as medical records collected over time may require many gigabytes of storage. Hence, a scalable alternative is to store a user's medical data on the application side. Although systems, like \sysName, can be deployed on application-owned private infrastructure, storing the data on the cloud forms a better choice due to the cloud's pay-per-use model and seamless scalability~\cite{journals/corr/RajanS13} (we discuss the challenges of hosting sensitive medical data on third-party cloud vendors below).

Moving on to the question of \textit{how} to store heterogeneous medical data, the users likely upload data in different formats -- PDFs, images, videos, CSV files, and so on. A medical data system must first process and parse all input records and provide a structure for the unstructured data. The system must next choose the type of underlying data storage model, such as relational, key-value, or columnar. Ideally, the system should use a hybrid data model to enable efficient data storage and retrieval. As an example, information about visits to a clinic can be stored in a relational model, e.g., SQL tables, whereas medical image data, such as X-rays, may need to be stored in unstructured formats, e.g., object stores. While we could treat this as a data lake problem~\cite{journals/pvldb/NargesianZMPA19,conf/sigmod/RamakrishnanSDK17,journals/pvldb/ArmbrustDPXZ0YM20,terrizzano2015data}, existing techniques are too generalized and as yet unclear how they will translate to medical records. Hence, research challenges here include studying the various formats that medical records typically come in and conducting a literature survey to identify the best data models to store these records. Moreover, beyond identifying promising data models for storage, an autonomous system should be able to parse any newly added medical records and automatically store the relevant information under relevant tables or data stores.

\resques{Identifying the best data models to store medical data and parsing new records to autonomously store them in relevant formats}



\begin{figure*}[!t]
\BBB\BBB
    \centering
    \includegraphics[scale=0.5]{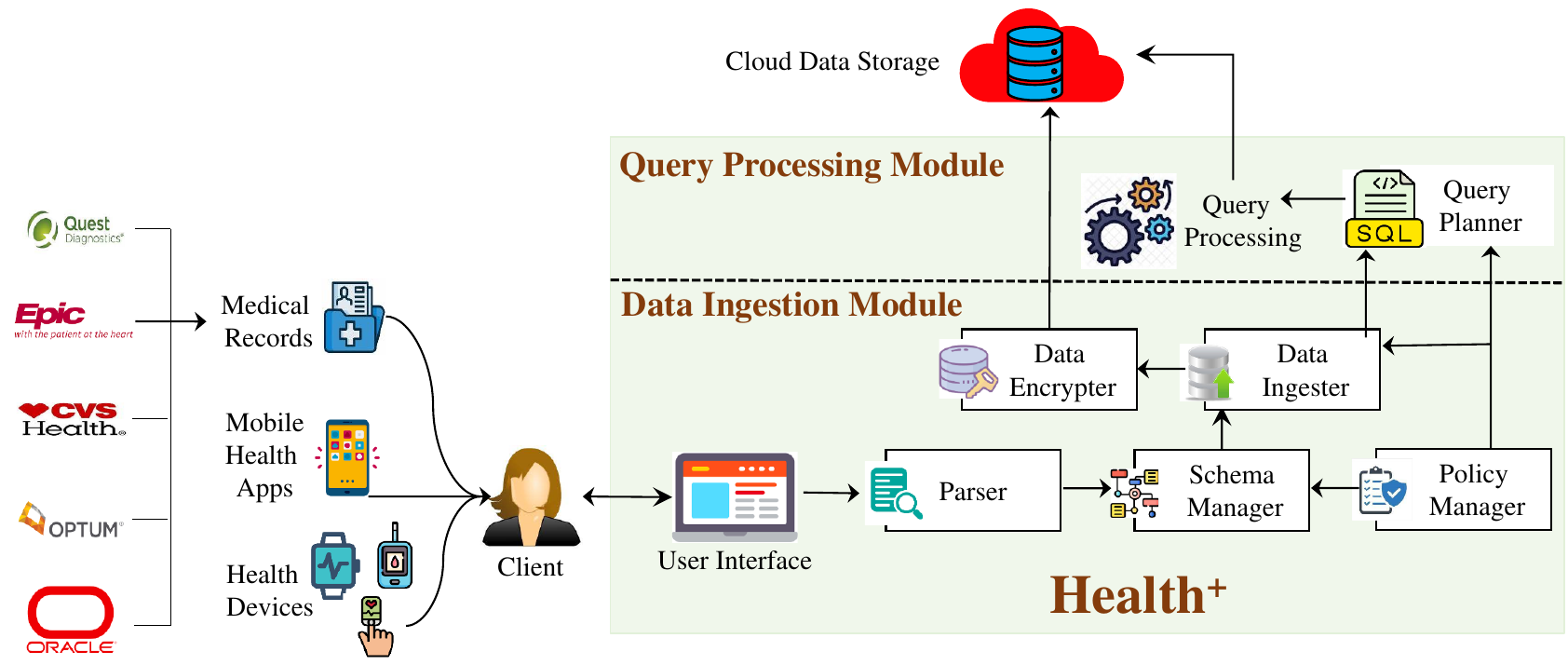}
    \BBB
    \BB\caption{\textsc{Health$^+$} Architecture.}
    \label{fig:model}
    \BBB
\end{figure*}

\medskip
\noindent\textbf{Data Fusion.}  

The data ingested into a health data management system, such as \sysName, may be fragmented. Suppose a user, Alice, suffers from disc herniation and visits a doctor, a lab technician, and a physiotherapist about the ailment. Alice uploads the doctor's notes, the laboratory results, and the physiotherapist's notes at different times, and the system may store them in different tables and formats. Later, she may decide to get a second opinion and wants to download all existing data about disc herniation. If a system is not well designed and encumbers its users with data retrieval, then Alice faces two main challenges here. First, she needs to remember all the data relevant to disc herniation. Second, she may need to possess technical skills, such as writing SQL queries to get the exact desired data. An alternative user experience is that the user inputs queries in plain language, e.g., ``retrieve records on to disc herniation,'' into the system and the system automatically fuses the fragmented data to give the final result. This experience reduces the burden on the user.

While automatic data fusion by the system sounds promising, achieving the functionality has technical challenges. First, the data relevant to an ailment may be stored in different formats, such as relational tables, object stores, and time series formats. Our system needs semantic intelligence to identify related data records to fuse them together. Next, not all data provided by a user has useful information, depending on the query posed by the user. For example, the time of visit to the doctors may not be useful when exporting data related to disc herniation. Next, when finding relevant records, data privacy is most important, disallowing the use of any third-party tools, like LLMs. Finally, similar to data augmentation, it is unclear when to fuse the data: fusing on the fly, i.e., \textit{after} a user queries the system, can be time-consuming, since the system needs to parse and combine a multitude of records, but fusing them a priori and storing them on a hard disk can lead to wasted storage. A robust health data system should address all the above challenges.

\resques{Efficient and accurate mechanisms to fuse related yet fragmented medical records}


\medskip
\noindent
\textbf{Data privacy and security.}

While outsourcing a health system's data to a third-party cloud has great potential in ensuring cost-effective scalability and high availability, storing highly sensitive data (such as medical records) on the cloud poses privacy and security threats. Storing such sensitive data in plaintext on cloud services is a non-viable solution because the cloud provider, who may not entirely be trustworthy, can intentionally (e.g., due to a rouge employee) or unintentionally (e.g., due to cyber-security attacks) leak individuals' ailments to malicious entities. Hence, systems such as \sysName must encrypt the medical data before outsourcing it to the cloud.

However, recent attacks have shown that data encryption is insufficient to preserve data privacy~\cite{islam2012access, grubbs2019learning,
demertzis2020seal}.S
These attacks exploit side-channel information such as the data access patterns (i.e., the identity of the objects satisfying a query) and the size/volume of results returned for queries to accurately uncover the underlying plaintext data. Existing systems that protect against such attacks 
(e.g.,~\cite{stefanov2013path, grubbs2020pancake, maiyya2023waffle, DBLP:conf/sigmod/AntonopoulosASE20}) 
cannot handle data stored in different formats, as is necessary for systems such as \sysName. Moreover, the health management system must decide how much trust to place on the cloud provider. If the system aims to tolerate malicious failures, wherein an adversary can tamper with the outsourced, it requires additional techniques to either enable detecting the data tampering (such as using techniques proposed in~\cite{haeberlen2007peerreview, yumerefendi2007strong,  jain2013trustworthy}) or ensure Byzantine fault tolerance by replicating the data across multiple servers (using Practical Byzantine Fault Tolerance~\cite{castro1999practical} technique). Hence, a secure health data management system must explore research directions to ensure the privacy and security of users' sensitive data.

\resques{End-to-end privacy and security of data, hiding side channel information, and ensuring tamper resistance}

\section{System Design}
\label{sec:system design}
\BB
This section presents a high-level overview of \sysName (see~{\color{blue}Figure~\ref{fig:model}}) and discusses its various components.

\noindent
\textbf{System model.}
Given our goal of empowering users/individuals (i.e., patients) to control and manage 
their medical data, they are at the center of \sysName, as shown in {\color{blue}Figure~\ref{fig:model}}. From the tools currently deployed by healthcare providers, users gather their medical records, which typically come in various formats, and upload them to \sysName. \sysName internally consists of a number of `intelligent' modules that process input data records to store them efficiently and execute queries on the stored data. We assume the deployer of \sysName owns a small fleet of trusted servers, which execute \sysName. However, the deployer may not possess enough infrastructure to host the large amounts of medical data uploaded by users. Hence, \sysName leverages the scalability, high availability, and cost-effectiveness of cloud-based storage services to store arbitrarily large medical data. Importantly, the users interact with the cloud-stored data by communicating with \sysName.

\medskip
\noindent
\textbf{Trust assumptions.}
We envision \sysName to be a trusted partner to the end users, wherein they
entrust \sysName with their data and queries. \sysName assumes healthcare providers to be trustworthy. In other words, healthcare providers do not provide any suspicious medical data to the user. Users are also considered trusted, and we assume that they upload their health records into \sysName without modifying them. The cloud is assumed to be untrusted due to potential data leaks from internal or external attacks, and hence, \sysName always encrypts the outsourced data to protect highly sensitive patient data. \sysName prevents revealing any information -- such as data in cleartext -- to the cloud.

\sysName consists of two modules, namely a \emph{data ingestion module} and \emph{a query module}, which are described below.

\begin{figure*}[!t]
    \centering
    \BBB\BBB
    \includegraphics[scale=0.12]{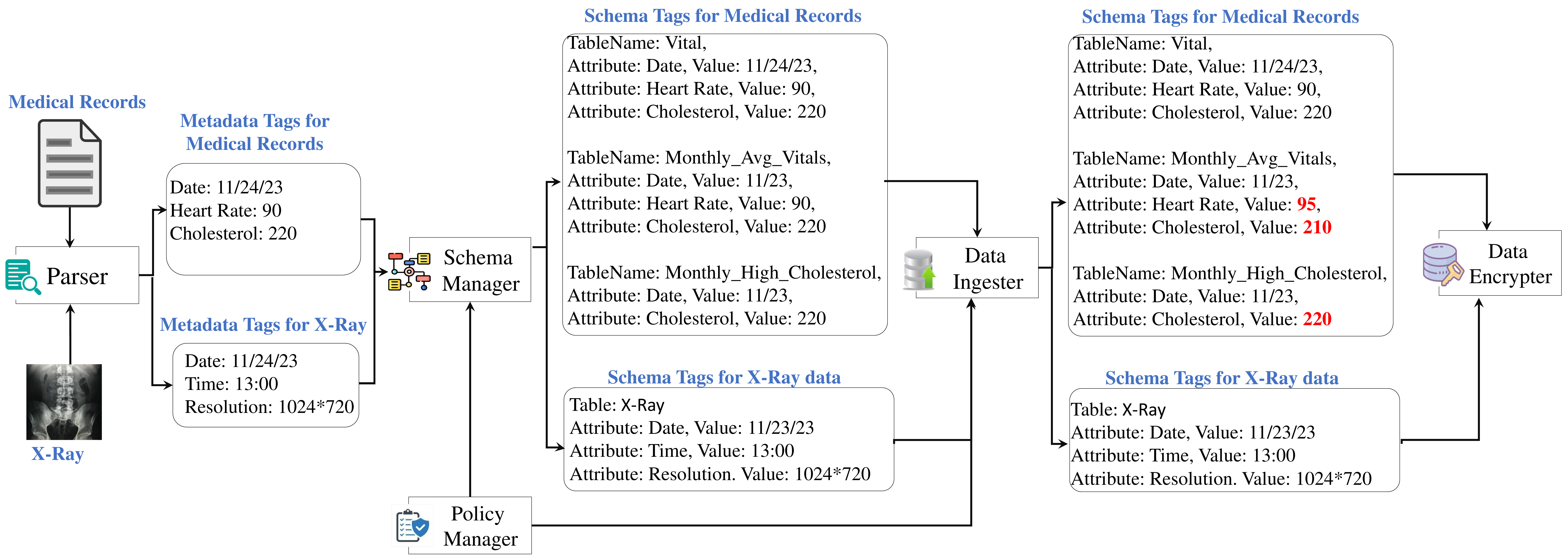}
    \caption{Dataflow in \textsc{Health$^+$} for a medical record with $\langle$Date: 11/24/23, Heart Rate: 90, Cholesterol: 220$\rangle$ and for an X-ray.  Inputs and outputs of several components of the data processing module are shown here. The red color represents the output of the data ingester, which is also shown in~{\color{blue}Example~\ref{sec:system design}.6}, after the update over the corresponding tables shown in~{\color{blue}Table~\ref{table:threetables}. }}
    \label{fig:dataflow}
\end{figure*}

\subsection{Data Ingestion Module}
\B
The data ingestion module enables users to upload their health data and store it securely at a public cloud. This module includes several components, as discussed below. Users primarily interact with their data by communicating with \sysName through the system's \emph{user interface}. The remaining components operate autonomously or with minimal user involvement, and are executed on the small yet trusted compute fleet owned by the agency that deploys \sysName. We explain the various components of the data ingestion module using an example, as illustrated in~{\color{blue}Figure~\ref{fig:dataflow}}.

\medskip
\noindent
\textbf{\ul{Graphical user interface (GUI)}} provides an easy option for individuals to upload their medical records and queries to be executed on the data through \sysName. Since users interact with most applications either via a mobile app or through a web app, \sysName aims to provide both interfaces for users. Users can upload data in several formats, e.g.,  
PDFs, 
images, movies, handwritten notes, or csv files (for data generated by smart devices). 
Many health apps also permit users to grant access to export data to other systems; e.g., Garmin~\cite{garmin} allows users to export the activities tracked by its watches to other apps, such as Strava~\cite{strava}, a sports-based social network. As advanced features, \sysName may allow for automatic synchronizations, where streams of health data (such as heart rate or sleep duration) from health apps can be directly uploaded to \sysName without requiring user intervention. Once the user uploads medical records, the parser processes it, as described next.

\medskip
\noindent
\textbf{\ul{Parser}}, as is indicative, parses the uploaded records, which can
come in many formats. Specifically, the parser leverages libraries such as \cite{pdf-parser1, pdf-parser2, image-parser, image-parser2} to parse and extract information from the files (such as PDFs and images) uploaded by the users.
The parser module then annotates the data with \emph{metadata tags} based on various categories, including reserved medical keywords, body mass index (BMI), heart rate, time, doctor's name, facility name, and others. In case, the concrete keywords cannot be extracted, the parser creates a generic `Description' keyword and populates it with the data provided in the input file. A metadata tag consists of two parts: the category keyword and its associated value in the data. Metadata tags, along with the entire uploaded data, then advance to the next component, which is a schema manager. For inputs from mobile health applications, since they can typically be modeled as time series data, the parser annotates the data by tagging time along with a keyword, such as $\langle$\textit{resting-heart-rate:time}$\rangle$ and its corresponding value.

\noindent
\textit{\textbf{Example~\ref{sec:system design}.1.}} {\color{blue}Figure~\ref{fig:dataflow}} shows that the parser receives two inputs: a PDF file and an X-ray image. Metadata tags for the X-ray image capture time, date, and resolution of the X-ray, while metadata tags for a PDF file include the reserved medical keywords (such as date, heart rate, and cholesterol).  $\blacksquare$


Although the output from the parser proceeds to the schema manager, we first explain the policy manager, since the schema manager requires information from both the parser and the policy manager ({\color{blue}Figure~\ref{fig:model}}).

\medskip
\noindent
\textbf{\ul{Policy Manager}} is responsible for defining and implementing \emph{storage policies}, \emph{enrichment policies}, and \emph{data sharing policies} based on the queries submitted by the users (not the data they upload).

\smallskip
\noindent
\textbf{\emph{1. Storage policies.}} 
One of the goals of \sysName is to enable efficient retrieval of data on commonly asked user queries. Our design mantra to achieve this goal is to store commonly asked queries as a separate table (except for images, videos, or other raw binary data, we assume \sysName stores its data in a relational model).
Hence, the policy manager automatically learns common query patterns over time and informs the schema layer on how to structure and store the data by generating storage policies. The invariant \sysName must always ensure that every input data resides in at least one table (or in raw format in an object store). 
As an advanced feature, \sysName can also accept user preferences and utilize them as policies. The storage policies indicate details to the schema manager, such as the number of tables to create, the schema for the tables, and whether any new tables should be added or some unused tables should be removed. It also ensures that every input data appears in at least one table (or in other storage formats). Moreover, the storage policies also dictate what indexes to create for each table based on frequently asked queries. These tasks relate to Research Directions~4 and~5.



\noindent
\emph{\textbf{Example~\ref{sec:system design}.2.}} Consider a commonly input data of \texttt{Vital}, as seen in {\color{blue}Table~\ref{table:threetables}(a)}. Each entry on \texttt{Vitals} consists of the date the vitals were measured and say heart rate and cholesterol levels. Assume that users typically query for the monthly average cholesterol and the monthly high cholesterol. While storing them all in a single table suffices to ensure accurate results for the two commonly asked queries, scanning through a single \texttt{Vitals} table and computing aggregates and maximum can be time-consuming, given that all our data is outsourced and stored in an encrypted format. Hence, the policy manager identifies the common query patterns and indicates the schema manager to create two additional tables, \texttt{Monthly\_Avg\_Vitals} (to store monthly average heart rate and cholesterol values) and \texttt{Monthly\_High\_Cholesterol} (to store a month's highest cholesterol value), with the necessary attributes (the data enricher module computes averages and maximums to fill in the necessary details automatically). 
Note that although we provide a simple example where new tables are created based on just a single \texttt{Vitals}, we envision this module to also create storage policies that span multiple tables and different storage formats. $\blacksquare$

\medskip
\noindent
\textbf{\emph{2. Enrichment policies.}}
Enrichment policies enable \sysName to construct missing values in the storage tables (such as aggregates on monthly average and high cholesterol) or extrapolate missing values in the input data itself, whenever possible (for the example discussed in Research Direction~3 where a user's vitals measured in one clinic can be linked to the details of a visit to another clinic on the same day, if the second clinic did not measure the vitals). One important decision the policy manager makes is when to execute the data enrichment process --- whether at data ingest time or at query processing time. This decision, again, depends on the query patterns and their frequencies. Since additional tables, such as \texttt{Monthly\_Avg\_Vitals} and \texttt{Monthly\_High\_Cholesterol}, were created in the last example to efficiently retrieve these values, enriching these tables (i.e., computing the aggregates) occurs at data ingest time. In contrast, for other less frequently queried tables having missing values, the enrichment occurs during query processing time. Whenever a policy enriches data as a form of extrapolation (due to missing information in the input data), \sysName clearly and explicitly marks it as an extrapolation and notifies the user.

\noindent
\textbf{\emph{Example~\ref{sec:system design}.3.}} For the two tables \texttt{Monthly\_Avg\_Vitals} and \texttt{Monthly\_High\_Cholesterol}, the policy manager creates a policy of $\langle$\textit{ingest\_time@}\texttt{Vital}$\rangle$, associated with the two aggregation tables. These policies indicate to the data enricher (explained later) that values for these tables should be computed and filled as soon as rows are added to \texttt{Vital} table. In contrast, consider a scenario where a user visits two clinics on the same day, whose details are stored in a table, say \texttt{Visit\_Details}, and only one of the two clinics measures vitals. Here, the policy manager creates a policy $\langle$\textit{process\_time@}\texttt{Vitals}$\rangle$, associated with  \texttt{Visit\_Details} table to indicate that when \texttt{Visit\_Details} is queried, some missing attributes might be found in the \texttt{Vitals} table, but only look for them at query processing time. $\blacksquare$

\begin{table}[!t]   
   \centering
    \begin{minipage}{.99\linewidth}
     
      \centering
        \begin{tabular}{|l|l|l|}\hline
            Date & Heart Rate & Cholesterol \\\hline
            10/1/23 & 90 & 190 \\\hline
            10/10/23 & 80 & 150 \\\hline
            11/5/23 & 100 & 200 \\\hline
            11/24/23 & {\cellcolor[HTML]{C1FD8E}{90}} & {\cellcolor[HTML]{C1FD8E}{220}} \\\hline
        \end{tabular}
         \subcaption{\texttt{Vital} table.}
    \end{minipage}%

    \begin{minipage}{.99\linewidth}
      \centering
        \begin{tabular}{|l|l|l|}\hline
            Month & Heart Rate & Cholesterol \\\hline
            10/23 & 85 & 170  \\\hline
            11/23 & {\cellcolor[HTML]{C1FD8E}{95}} & {\cellcolor[HTML]{C1FD8E}{210}} \\\hline
        \end{tabular}
      \subcaption{\texttt{Monthly\_Avg\_Vitals}.}
    \end{minipage} 
    
        \begin{minipage}{.99\linewidth}
      \centering
    \begin{tabular}{|l|l|}\hline
            Date & Cholesterol \\\hline
            10/23 & {190} \\\hline
            11/23 & {\cellcolor[HTML]{C1FD8E}{220}} \\\hline
        \end{tabular}
        \subcaption{\texttt{Monthly\_High\_Cholesterol}.}
    \end{minipage} 
     \caption{Three different tables. The green color represents a new row in \texttt{Vital} table and the resulting updates in the other two tables. Dataflow among different components of the data processing module due to this new row is shown in~{\color{blue}Figure~\ref{fig:dataflow}}.}
    \label{table:threetables}
    \BBB
\end{table}

\medskip
\noindent
\textbf{\emph{3. Data sharing policies.}} Data sharing policies, which relate to Research Direction 2, enable a user to fetch only appropriate records to be shared during query execution time. The primary use-case is when users want to share medical records about past diagnoses and/or treatment during a new diagnosis. Sharing all past medical records can impact a user's privacy (please refer to the example in~{\color{blue}\S\ref{sec:research-dirs}} Research Direction 2), but it is essential to share only the necessary records for the current diagnosis. To achieve this, we envision deploying federated learning~\cite{journals/npjdm/RiekeH0MRABGLMO20,dun2023efficient} to train a model on associations between records and current diagnoses. These policies take effect during query processing, where, based on a user query, the query processing module retrieves all necessary documents and information that the user can present to the healthcare professional.

\noindent\textit{\textbf{Example~\ref{sec:system design}.4.}} Consider a patient suffering from both disc herniation and OCD. The policy manager would create a data sharing policy that links disc herniation with tables and documents such as all MRIs and X-rays of discs, related medications, and physical therapy plans, but with no links to diagnosis and treatments for OCD. $\blacksquare$

Having discussed the policy manager, which affects both data storage and query processing, we continue our discussion on the next steps that occur during data ingestion.

\medskip
\noindent
\textbf{\ul{Schema manager}} is responsible for the following three tasks: creating tables (if tables do not exist), performing entity resolution, and deciding how to store the data in tables by producing schema tags. In general, the schema manager maintains information about all the tables and other storage formats used in \sysName.

\noindent
\textbf{\emph{1. Table creation.}} The schema manager receives all the parsed data (i.e., metadata tags) from the parser. The schema manager then consults the policy manager to learn the storage policies that relate to the input data. Based on the policies, it creates new tables if such tables do not already exist. If all tables related to the input data exist, then the schema manager advances to the next step.


\noindent
\textbf{\emph{2. Entity Resolution.}} The schema manager performs entity resolution over the metadata tags~\cite{conf/icde/SancaA23}. 
Let us consider the following example: an individual visits two different doctors, and they provide information about their body mass index (BMI). Here, the parser may have created two different metadata tags: 
$\langle$Body Mass Index, 30$\rangle$ and $\langle$BMI, 30$\rangle$, based on the exact keyword-texts used in the medical record. Since both values refer to the same measure, the schema manager performs entity resolution and updates both metadata tags to, for example, BMI.

\noindent
\textbf{\emph{3. Schema Tags.}} Next, the schema manager, based on the metadata tags, 
produces \emph{schema tags}, to indicate to the next component (i.e., data ingester), the attribute values to be stored in appropriate tables. Notice, neither metadata tags nor schema tags reside in storage; their primary roles lie in accurately
transforming the unstructured input data to structured and store them appropriately.
Given that metadata tags already contain all the keywords and their values provided
in the input data, schema tags add 
which table to store those key-value pairs in and under what column.

\noindent
\emph{\textbf{Example~\ref{sec:system design}.5.}} We explain the concept of schema tag using an example.
Say a PDF uploaded by a user contains details on their heart rate and cholesterol values measured on a specific date. The parser would have created three metadata tags
with date, heart rate, and cholesterol as keys and their corresponding values. 
The schema manager consumes the metadata tags and creates schema tags. Consider the three tables created  by the schema manager:  
\texttt{Vital}, \texttt{Monthly\_Avg\_Vitals}, and \texttt{Monthly\_High\_Cholesterol};~{\color{blue}Table~\ref{table:threetables}}. 
Suppose, the schema manager receives the following metadata tag: \texttt{$\langle$Date: 11/24/23, Heart Rate:  90, Cholesterol: 220$\rangle$};~{\color{blue}Figure~\ref{fig:dataflow}}.
For this metadata tag, the schema manager produces the following three schema tags: 

\begin{enumerate}[noitemsep,nolistsep,
leftmargin=0.01in]
\item
\texttt{$\langle$TableName: Vital, 
Attribute: Date, Value: 11/24/23, Attribute: Heart Rate, Value: 90, Attribute: Cholesterol, Value: 220$\rangle$}

\item
\texttt{$\langle$TableName: Monthly\_Avg\_Vitals, Attribute: Date, Value: 11/23, Attribute: Heart Rate, Value: 90, Attribute: Cholesterol, Value: 220$\rangle$}

\item
\texttt{$\langle$TableName:~Monthly\_High\_Cholesterol, Attribute: Date, Value:~11/23, Attribute: Cholesterol, Value:~220$\rangle$}

\end{enumerate}

Notice, schema tags for \texttt{Monthly\_High\_Cholesterol} do not contain details on heart rate. The data enricher module, which we describe next, will compute the aggregates and store those values; not the values produced in the schema tags. Schema tags mainly indicate the tables and their columns that will
be affected by newly uploaded documents, and the attribute values depend on other policies.






We next discuss how \sysName enriches and ingests the data into the cloud storage.

\medskip
\noindent
\textbf{\ul{Data enricher}} is responsible for enriching the data and also creating or
maintaining indexes. This component receives schema tags from the
schema manager and consults the policy manager for enrichment policies. Based on that,
it identifies what tables and columns need to be populated as part of enrichment
-- either to store some pre-computed aggregate data or to extrapolate data missing
in the input -- and executes the necessary steps to fill in the missing data. Computing
aggregates such as average or max of an attribute may require fetching data from
the cloud storage or depending on the indexes, might be computed locally. 

\noindent\textit{\textbf{Example~\ref{sec:system design}.6.}} Based on~{\color{blue}Example~\ref{sec:system design}.3} provided in the policy manager,
tables \texttt{Monthly\_Avg\_Vitals} and \texttt{Monthly\_High\_Cholesterol}
have an associated enrichment tag $\langle$\textit{ingest\_time@}\texttt{Vitals}$\rangle$. 
On receiving schema tags,~{\color{blue}Example~\ref{sec:system design}.5}, 
the data enricher computes the monthly average and monthly maximum values (based on schema tags 2 and 3 of~{\color{blue}Example~\ref{sec:system design}.5}) and updates the attribute values in the schema tags, as follows:

\begin{center}
\footnotesize
\texttt{$\langle$TableName: Monthly\_Avg\_Vitals, Attribute: Date, Value: 11/23, Attribute: Heart Rate, Value: 95, Attribute: Cholesterol, Value: 210$\rangle$}

\texttt{$\langle$TableName: Monthly\_High\_Cholesterol, Attribute: Date, Value: 11/23, Attribute: Cholesterol, Value: 220$\rangle$} $\blacksquare$
\end{center}

\noindent
\textbf{\emph{Indexes.}} We consider creating and maintaining indexes as also a form of
data enrichment, and hence, the enricher is tasked with this responsibility as well.
The policy manager provides index-related information in storage policies, which is 
leveraged by the data ingester to initially create and then keep the indexes up-to-date. 
The data enricher utilizes all the schema tags to create and maintain indexes locally (i.e., indexes are not
outsourced to the cloud storage).



 
\medskip
\noindent
\textbf{\ul{Data encrypter}} encrypts the data and outsources them to the cloud. This is the last step of the data ingestion module. Particularly, the encrypter inherits all the schema tags populated over the data enricher
and implements cryptographic techniques on them and places in the desired tables, which are outsourced to the cloud. 
As indicated in~{\color{blue}\S\ref{sec:research-dirs}} Research Direction 6, data encryption may itself be insufficient to ensure complete privacy of the outsourced medical records due to the side-channel information, such as access patterns that an adversary can exploit.
However, executing techniques that ensure stronger privacy, such as Oblivious RAM-based 
storage~\cite{stefanov2013path,ren2015constants} or secure multi-party computation based
storage~\cite{DBLP:journals/cacm/Lindell21}, incurs extremely high latency overheads.
Hence, in the first iteration, we aim to build \sysName by preserving data confidentiality through encryption, and then to add techniques to also hide all side-channel leakages iteratively. To enable higher functionality that the server can execute, we aim to encrypt data using deterministic and order-preserving encryption schemes~\cite{DBLP:conf/sigmod/AgrawalKSX04}. This allows the server
to execute point and range queries on encrypted data without having to communicate with the trusted component of \sysName.


\subsection{Query Processing Module}
\B
This section describes how \sysName executes queries. 
Users utilize the GUI of \sysName to query their medical data. Since we assume the users of \sysName to be non-technical individuals, they can input their queries in plain language, e.g., `what was my maximum cholesterol in November 2023.' 
To enable such functionality, we aim to employ large language models~\cite{journals/pvldb/FernandezEFKT23} that can translate plain language queries into machine-understandable queries, such as SQL. After constructing appropriate queries, such as in SQL, the query processing module generates an optimized query plan similar to existing query optimizers~\cite{conf/sigmod/BegoliCHML18}. 
Note that all the metadata necessary to optimize queries will be stored in the
trusted component of \sysName, and not on the cloud. Once the query processor generates an optimized query execution plan, it encrypts all the predicates using the
appropriate encryption mechanism (i.e., deterministic or order-preserving) such that the cloud storage service can execute the query to retrieve the results.

\sysName consists of two unique query processing features: (i) supporting data enrichment at query time, and (ii) enabling data sharing that limits the shared records to only what is necessary in a given situation. Again, since the policy manager would have created enrichment and data sharing policies, the query processor only needs to consult the policy manager to retrieve the appropriate policy based on the user's query and take the necessary action. For example, as described in the policy manager section,  \texttt{Visit\_Details} table will have an associated policy $\langle$\textit{process\_time@}\texttt{Vitals}$\rangle$. When a user's query reads data from  \texttt{Visit\_Details} table, if the retrieved rows have missing values on vitals, the query processor fetches these details from  \texttt{Vital} table.
Similarly, when a user's query asks to share all details related to say disc herniation, the query processor consults the policy manager's data sharing policies and accordingly retrieves records on MRIs and X-rays of discs, related medications, and physical therapy plans, but nothing more. If these records are insufficient, users
can also send another query based on their interactions with the healthcare 
professional.

\section{Expanding Horizons}
The above sections have outlined the design of \sysName, yet, there are the following other directions to be followed:

\medskip
\noindent
\textbf{\textsc{Health$^+$} at the cloud.} 
We initially assumed that the two modules, namely data ingestion and query processing, in \textsc{Health$^+$} would be executed at a trusted entity. We envision that all the components of \textsc{Health$^+$} be executed at the public cloud. This transition necessitates secure data processing by all \textsc{Health$^+$} components, opening up intriguing avenues in secure data processing.
For example, at the data ingestion module,  the schema manager needs to operate over the encrypted metadata tags and encrypted policies to produce schema tags for data enricher. A possible solution is to use secure hardware (e.g., Intel Software Guard eXtension (SGX)~\cite{costan2016intel}). However, SGX is prone to side-channel attacks, e.g., cache, page-fault attacks~\cite{brasser2017software,shih2017t}. This demands a meticulous design and development of algorithms over SGX for each \textsc{Health$^+$} component. Another, more substantial solution that could unite the crypto and hardware communities is to process as much as possible over ciphertext data and minimize processing within SGX. For instance, executing queries entirely on encrypted data using cryptographic techniques, while limiting the data processing module to SGX.

\medskip
\noindent
\textbf{Malicious entities.} While healthcare providers are always trusted, the data received from them is vulnerable to corruption, either accidentally or intentionally by individuals with motives such as legal action against a doctor. In other words, the \emph{user can behave maliciously}. To protect against the storage of inaccurate data in the system, it is crucial to verify the data by comparing it against the records provided by healthcare providers and utilizing tamper-proof evidence. A straightforward solution to address this issue is to have the healthcare provider verify the entire dataset before storing it in the cloud via \sysName. However, this approach poses a substantial burden on healthcare providers and does not offer a mechanism to detect future alterations to the data by users.
To mitigate this risk, we can employ a decentralized system to store the hash digest of the data. Specifically, healthcare providers can compute the hash digest over the data and store the digest in a decentralized database (or a blockchain system). Upon receiving data (or any updates to existing data) from the user, the parser of \sysName can similarly compute the hash digest, store it in the decentralized database, and verify it against the digest recorded by the healthcare providers.

The \emph{public cloud can also be malicious}, potentially altering the encrypted data or providing incorrect or incomplete answers to the user. Existing works on authenticated data structures~\cite{miller2014authenticated}, zero-knowledge proofs~\cite{DBLP:conf/stoc/GoldwasserMR85}, and result verification~\cite{DBLP:journals/pvldb/YueZZCLO23} could be applied by \sysName to verify the results of SQL queries. However, these techniques are not explicitly designed to verify the work of data enrichment, a crucial step in producing the final answer to the query. Additionally, such techniques typically require appending extra verifiable information during data ingestion. In the case of \sysName, which supports a wide variety of dynamic operations, including dynamic policies, it becomes challenging to append verifiable information to the data. This complexity arises from the continuous changes in data, tables, and results over time, influenced by storage, enrichment, and sharing policies.

\medskip
\noindent
\textbf{Enhancing Security.} 
Secret-sharing techniques~\cite{shamir1979share} provide the utmost data security by splitting a secret into multiple pieces and distributing them across non-colluding cloud servers. Despite their robust security features, these techniques may experience efficiency challenges when applied to large datasets. However, since an individual's medical data, accumulated over their lifetime, is typically not massive, there are potential avenues for exploration. Firstly, there is potential for modifying these techniques for on-the-fly data enrichment and query execution. Secondly, addressing common leakages related to access patterns and volume during query execution from secret-sharing techniques is of paramount importance. Methods can be devised to selectively disclose partial access patterns and volume, guided by user preferences or data sensitivity. This can be achieved by creating data buckets (similar to hash buckets~\cite{shapiro1986join} or data bucketization~\cite{hacigumucs2002executing}) that are processed to derive the final data. Additionally, enhancing support for efficient indexing methods over secret-shared data is a noteworthy area of exploration. Techniques that combine indexes for secret-sharing, such as S3ORAM~\cite{hoang2017s3oram}, on specific data segments and utilize linear scans over the remaining data could be developed.

\medskip
\noindent
\textbf{Interconnection of existing medical systems and \textsc{Health$^+$}.} The proposed \sysName operates on the assumption that users will obtain data from healthcare providers and manually input it into the system. While this method reduces the burden on healthcare providers, it may pose challenges for users unfamiliar with computer/mobile systems, impeding their ability to upload data. Ideally, healthcare providers should take charge of inserting user data into \sysName. This integration process must be seamless for healthcare providers, avoiding the necessity of adopting a new medical system. The replacement of the existing medical system (e.g., Epic) with a new one presents challenges in terms of development and political issues in various countries. Hence, a preferable solution is to provide healthcare providers with the option to utilize their existing systems for recording user medical data, with the data concurrently and seamlessly added to \sysName without requiring direct intervention from healthcare providers. This approach not only eases challenges for users but also enhances the system's capability to manage potential malicious users, as discussed in the first direction in this section.

\medskip
\noindent
\textbf{Collaborative analysis, logging, and auditing.} 
Medical data serves various purposes for research agencies, including drug discovery and early disease detection, through anonymous data sharing. However, due to substantial privacy challenges associated with medical data sharing, numerous healthcare providers refrain from sharing their data, even in critical situations~\cite{galaitsi2021challenges,dron2022data}. For instance, during the COVID-19 pandemic, several hospitals were unable to share data with the Department of Health \& Human Services (HHS)~\cite{healthcare-covid19-data-1, healthcare-covid19-data-2, dron2022data}. Even when attempts are made to share data, research agencies encounter challenges in having a unified view of the data. 

\sysName aims to address this issue by keeping the entire medical data of an individual, including data from medical apps/devices, in a single location. However, this approach introduces new challenges. These include determining how data will be shared with research agencies, devising effective anonymization methods, and establishing mechanisms for users to express their preferences regarding data sharing. Additionally, legal laws such as Health Insurance Portability and Accountability Act (HIPAA)~\cite{hippa} pose an additional challenge in the form of tamper-proof logging. This involves logging queries by the user or research agencies to understand their requests and whether the system aligns with the underlying sharing policies. Privacy-preserving auditing of logs by third parties further complicates matters, as these entities are not permitted to access an individual's medical data but seek to verify operations conducted on the data against established policies. Recent solutions, such as SMCQL~\cite{DBLP:journals/pvldb/BaterEEGKR17}, have attempted to develop cryptographic solutions for medical data sharing; however, they do not consider policy-based data sharing, logging, and auditing. Also, they suffer from inefficiency during data sharing.

\end{document}